# A method for optimizing the structure of the software and hardware complex of a distributed process control system for large industrial enterprises

1st Ruslan Zakirzyanov
*NEXT engineering, LLC*
Kazan, Russia
zr@nexteng.ru

*Abstract*—The article proposes a method for optimizing the structure of the software and hardware complex of an automated control system for continuous technological processes for large industrial enterprises. General information is given on the relevance of the problem of choosing the structure of a system built on the basis of serially produced components, a formal description of the optimization problem is given, the criterion and limitations are highlighted. A solution method using the metaheuristic algorithm of ant colonies is described. A numerical example of the solution is given, the results of the algorithm are analyzed, and directions for further research are determined.

## I. Introduction

Modern process control systems for large industrial enterprises are usually built on the basis of software and hardware complexes [1], the components of which are mass-produced by industry. The properties and functionality of such distributed control systems are largely determined by the structure of the hardware complex, since the characteristics of the components used are specified by the manufacturers and are not subject to change. Thus, to achieve the required system characteristics, it is necessary to build a structure from components with known characteristics that would be as minimal as possible in cost and satisfy all requirements and restrictions, i.e. would be optimal. Building an optimal structure is an important task that is currently mainly solved empirically based on the experience of the designer and recommendations of the manufacturers of software and hardware. The task becomes especially important when designing control systems for large hazardous geographically distributed facilities with a large number of parameters. This article considers the formalization of the problem of optimizing the structure of the software and hardware complex of an automated process control system, proposes a solution to it, and provides recommendations for further development of research in this area.

## II. Features of distributed control systems for continuous technological processes

Technological processes at industrial enterprises are conventionally divided into continuous and discrete [2]. Continuous technological processes are characteristic of the oil and gas, chemical, petrochemical industries, and also partially metallurgy and thermal power engineering. Control objects of large industrial enterprises of the specified industries have their own characteristic features. Such objects, as a rule, are territorially distributed, are dangerous and technically complex, and have continuous processes. Changes often occur at such objects: the parameters of the technological mode change, some equipment is taken out for repair. Commissioning of such objects occurs in parts or stages. The software and hardware complexes mass-produced by industry for building on their basis an automated process control system for large industrial enterprises with continuous technological processes are called DCS – Distributed Control System in the literature [3]. In this article, "DCS" is understood exclusively as a specialized software and hardware complex for building control systems for complex continuous technological processes.

When creating a system, the designer of the automated process control system solves a complex problem. It is necessary to select such components for the system and build such an architecture from them so that the task of controlling the object is solved qualitatively, the system meets reliability requirements and is minimal in cost.

## III. Description of the structure

To formulate the problem of finding the optimal structure of the DCS, we will represent its structure as a tree (acyclic graph), an example of which is shown in Fig. 1. The number of levels in the structure is specified by the designer and can vary.

Methods for optimizing hierarchical structures are discussed in detail in [4, 5, 6].

Let the hierarchical structure of a distributed control system be given in the form of a tree $\mathcal{G} = (\mathcal{V}, \mathcal{E})$, where $v \in \mathcal{V}$ are devices (graph nodes), $\mathcal{E}$ are graph edges (communication channels between devices).

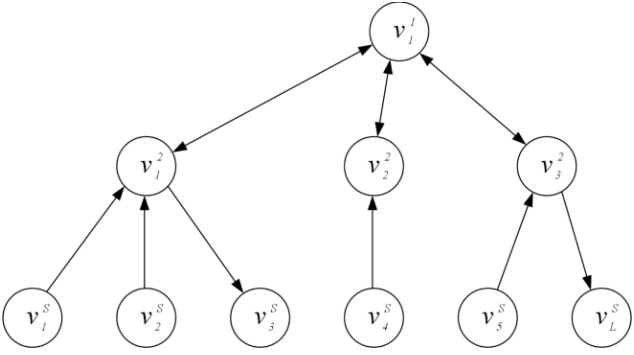

Fig. 1. Hierarchical structure of the DCS

Let the set of device types $\mathcal{U} = \{u_1, \dots, u_U\}$, consisting of $U$ device types, be given.

Let us assume that each node of the structure (device) of any type performs the same type of actions, consisting of three phases of the operating cycle of the device's internal program:

- Collecting (reading) information from the control object, or from nodes of the previous level of the hierarchy.
- Processing information (implementation of control algorithms).
- Issuing information (writing) to a lower level, or influencing the control object.

A device that has the function of processing information will be called a processor. A device that can only transmit information will be called a repeater.

Each type $u_i \in \mathcal{U}$ of a structure node is characterized by the following parameters:

- Cost of the device $C_i \in \mathbb{R}^+$.
- Number of connected physical channels $N_i \in \mathbb{N}$.
- Maximum memory capacity $R_i \in \mathbb{R}^+$.
- Probability of device failure $P_i \in [0,1]$.
- Performance (execution time of one program instruction) $T_i \in \mathbb{R}^+$.
- Operating mode $y_i \in \{0,1\}$ (1 – processor, 2 – repeater).
- Maximum number of child devices $M_i \in \mathbb{N}$.
- Transmission delay for repeaters $\tau_i \in \mathbb{R}^+$.

Each device type is represented by a vector:

$$u_i = (C_i, N_i, R_i, P_i, T_i, y_i, M_i, \tau_i), \quad i = \overline{1,U}. \tag{1}$$

Let a set of control loops $\mathcal{A} = \{a_1, \dots, a_A\}$, also be given, consisting of $A$ loops $a_j \in \mathcal{A}$:

$$a_j = (n_j, r_j, w_j), \quad j = \overline{1,J}, \tag{2}$$

where $n_j \in \mathbb{N}$ is the number of physical signals in the circuit, $r_j \in \mathbb{R}^+$ is the amount of memory required to store all instructions and variables of the circuit, $w_j \in \mathbb{N}$ is the number of instructions in the signal processing program of this circuit.

For this problem, we assume that all "field" signals are distributed among the loops in advance, the number of signals in each loop is known, the number of instructions in the loop processing program is known, the amount of memory required to store all loop variables and all instructions of the loop processing program is also known (estimated) in advance. We assume that the same time $T_i$ is required to execute any instruction of the loop processing program.

The structure of the optimal control system is a hierarchical structure (tree) with the number of levels $s = S \in \mathbb{N}$, where $s = 1$ corresponds to the root of the tree, i.e. the top of the hierarchy, $s = S$ corresponds to the level of tree leaves, i.e. the lowest level of the hierarchy. $L \in \mathbb{N}$ is the number of tree leaves, and the set of tree leaves is denoted by $\mathcal{L} \subseteq \mathcal{V}$. $V \in \mathbb{N}$ is the total number of devices in the structure. $K_s$ is the number of devices at level $s$. The number of tree leaves $K_S = L$. The set of leaves $\mathcal{L} = \{v_1^S, \dots, v_L^S\}$.

Based on the above, for each device $v_k^s \in \mathcal{V}$ we introduce the following functions:

- $u(v_k^s) = u_k^s \in \mathcal{U}$ – device type.
- $\pi(v_k^s) \in \mathcal{V} \cup \{\emptyset\}$ – parent device (parent node).
- $\mathcal{D}(v_k^s) \subseteq \mathcal{V}$ – set of child nodes.
- $\mathcal{D}'(v_k^s) \subseteq \mathcal{V}$ – subtree of node $v_k^s$, such that $\mathcal{D}(v_k^s) \subseteq \mathcal{D}'(v_k^s)$.
- $\mathcal{P}(v_k^s) \subseteq \mathcal{V}$ – path from node $v_k^s$ to the root of the tree.

Here $s \in \{1, \dots, S\}$ is the level in the hierarchy, $k \in \mathbb{N}$ is the serial number of the node at this level. Thus, each device can also be represented as a vector (3):

$$v_k^s = (u_k^s, \pi(v_k^s), \mathcal{D}(v_k^s), \mathcal{D}'(v_k^s)), \quad s = \overline{1,S}, \quad k = \overline{1,K_s}. \tag{3}$$

To solve the problem, we assume that only devices of level $S$ (leaves) are directly connected to the field equipment of the process plant, i.e. they can receive information from the control object and issue control actions to implement proper control of the object. We assume that there are no horizontal links between devices. Each circuit must be physically connected to some leaf of the tree. Leaf $v$ can itself process the circuits connected to it if it is a processor ($y = 1$), or pass the circuit for processing to a higher processor if the leaf is a repeater. Each circuit must be assigned to a single processor. A processor can only process circuits connected to leaves from its subtree. We will assume that only repeaters introduce a delay. Processors cyclically execute programs for processing all circuits assigned to them. The operating time of a processor is equal to the total processing time of all circuits assigned to it. Let us also introduce additional variables (the indices $s$, $k$ and $j$ are omitted below to simplify notation).

- $x_{va} \in \{0,1\}$, where $x_{va} = 1$ if the circuit $a$ is connected to the leaf $v$ ($\forall v \in \mathcal{L}, \forall a \in \mathcal{A}$), otherwise $x_{va} = 0$;
- $z_{va} \in \{0,1\}$, where $z_{va} = 1$ if the circuit $a$ is assigned to the processor $v$ ($\forall v \in \mathcal{V}, \forall a \in \mathcal{A}$), otherwise $z_{va} = 0$.

We will use the following terminology below. If the control loop is physically connected to a node of level $S$, we will say that loop $a$ is connected to leaf $v$ ($x_{va} = 1$). If the signals of loop $a$ are processed in node $v$, we will say that loop $a$ is assigned to node $v$ ($z_{va} = 1$).

## IV. RESTRICTIONS

For a tree-like structure of a distributed control system, the following restrictions are valid. The tree has a single root (4). The root of the tree can be either a processor or a repeater.

$$\exists! v_1^1 \in \mathcal{V}: \pi(v_1^1) = \emptyset. \qquad (4)$$

Nodes located at level $S$ (leaves of the tree) cannot have children:

$$|\mathcal{D}(v)| = 0, \quad \forall v \in \mathcal{L}. \qquad (5)$$

Each node can have a limited number of child devices (6). For example, a network switch or controller may have a limited number of Ethernet ports.

$$|\mathcal{D}(v)| \leq M_i, \quad \forall v \in \mathcal{V}\setminus\mathcal{L}, \qquad (6)$$

where $u(v) = u_i$. No more than $M_i$ child nodes can be connected to a node $v$ of type $u_i$. Expression (6) is valid for all nodes except leaves ($\mathcal{L}$).

Each circuit can only be connected to one sheet:

$$\sum_{x \in \mathcal{L}} x_{va} = 1, \quad \forall a \in \mathcal{L}. \qquad (7)$$

Let us write down the restrictions that characterize the processing of control loops by processor nodes. Only processors can process loop signals, i.e. a loop cannot be assigned to a node that is not a processor:

$$z_{va} \leq y_v, \quad \forall v \in \mathcal{V}, \quad \forall a \in \mathcal{A}. \qquad (8)$$

There can only be one processor for each loop:

$$\sum_{v \in \mathcal{V}} z_{va} = 1, \quad \forall a \in \mathcal{A}. \qquad (9)$$

Each processor has a memory limit (10). If the total memory required to store information for all the circuits assigned to a given processor exceeds the memory size of the given node type, the node will not be able to process all the circuits. In this case, an additional processor is needed to take on some of the load.

$$\sum_{a \in \mathcal{A}} z_{va} r_a \leq R_v, \quad \forall v \in \mathcal{V}. \qquad (10)$$

A processor can only handle paths connected to leaves of the processor's subtree (11). Thus, a processor only services its own subtree; paths connected to leaves of other subtrees cannot be assigned to this processor.

$$z_{va} \leq \sum_{\substack{v' \in \mathcal{L} \\ v' \in \mathcal{D}'(v)}} x_{v'a}, \quad \forall v \in \mathcal{V}, \quad \forall a \in \mathcal{A}. \qquad (11)$$

The total number of signals of all circuits connected to the $v$ sheet must not be greater than the number of channels of this sheet (12). The number of channels of the $N_v$ sheet is determined by the node type. The processor may have $N_v = 0$.

$$\sum_{a \in \mathcal{A}} x_{va} n_a \leq N_v, \quad \forall v \in \mathcal{V}. \qquad (12)$$

Since the designed structure is the structure of a control system for a dynamic object, it is important to take into account the dynamic properties of the system during synthesis. In practice, this is a rather complex task [7]. As an indicator characterizing the quality of the control system, we will take the program cycle time (quantization period) for one control loop. For any loop, regardless of its connection and purpose, the processing time $T_{cont}$ should not exceed the specified time $T_{max} \in \mathbb{R}^+$. The loop processing time consists of the processing time of all instructions of all loops assigned to a given node and the total time of all delays introduced by repeaters on the way from the leaf to the processor (13).

$$T_{cont} = \max_{a \in \mathcal{A}}\left[\sum_{a' \in \mathcal{A}} z_{v(a)a'} \cdot w_{a'} \cdot T_{v(a)} + \sum_{v' \in \mathcal{P}(v(a),a)}(1 - y_{v'})\tau_{v'}\right] \leq T_{max}, \qquad (13)$$

where $v(a)$ is the processor of loop $a$, $\mathcal{P}(v(a), a)$ is the path from the leaf with loop $a$ to the processor $v(a)$, $(1 - y_{v'})$ is the repeater indicator (1 for repeaters and 0 for processors).

We use the probability of failure as an indicator of system reliability. We consider all failures of devices in the system to be independent. The probability of system failure $P_{sys}$ in this case will be equal to the product of the failure probabilities of all its components. The methodology for calculating the reliability of hierarchical automated process control systems is given in [8]. The probability of system failure should not exceed the specified value $P_{max} \in [0,1]$:

$$P_{sys} = 1 - \prod_{v \in \mathcal{V}}(1 - P_v) \leq P_{max}, \qquad (14)$$

where $P_v$ is the probability of failure of the $v$-th device. We assume that $P_v = const$. In real systems, the probability of failure is a function of time: $P_v = f(t)$ [9].

We will also write down the conditions for coordinating processors. A processor cannot have a higher processor:

$$y_v + y_{\pi(v)} \leq 1, \quad \forall v \neq v_1^1. \qquad (15)$$

A processor cannot have processors in its subtree:

$$y_v + \sum_{v' \in \mathcal{D}'(v)} y_{v'} = 1, \quad \forall v \in \mathcal{V} : y_v = 1. \quad (16)$$

There must be exactly one processor on the path from any leaf to the root:

$$\sum_{v' \in \mathcal{P}(v)} y_{v'} = 1, \quad \forall v \in \mathcal{L}, \quad (17)$$

where $\mathcal{P}(v)$ is the path from leaf v to the root (including v and the root).

## V. TARGET FUNCTION

Previously [10, 11] the key criteria for optimizing the hierarchical structure of a distributed control system were defined.

The optimal hierarchical structure $\mathcal{G}_O$ should minimize the cost of the system under the restrictions specified in the previous paragraph. Thus, the optimal structure can be determined as a result of solving the optimization problem:

$$C_O = \min_{\mathcal{G}} \sum_{v \in \mathcal{V}} C_v, \quad (18)$$

where $C_O$ is the optimal total cost of creating and operating the system.

Thus, the structural optimization problem is formulated as follows. It is necessary to select such a hierarchical structure from nodes of predetermined types in order to deliver the minimum of the objective function (18) under the given constraints (4)…(17). We assume that an unlimited number of devices of each type can be used. We also assume that the desired structure is a homogeneous tree, and the path from any leaf to the root always contains the same number of nodes.

## VI. CHOICE OF METHOD AND SOFTWARE IMPLEMENTATION

For such large-scale problems, heuristic and metaheuristic algorithms can be used, such as the genetic algorithm and its modifications, the ant colony algorithm, particle swarm, simulated annealing, etc. The ant colony algorithm was chosen to solve this problem since it encodes the problem represented as a graph quite well, unlike genetic algorithms that are focused on systems with a constant structure. The ant colony algorithm and its practical application options are described in detail in [12, 13], a comparison of the genetic algorithm and the ant colony algorithm is given in [14].

To implement the algorithm, a Python program was developed that is capable of solving numerical examples with sufficiently large dimensions. The initial data for the numerical example are given in Table I. The example considers the case when there is a set of $A$ identical loops with parameters $n_1 = 1$, $r_1 = 1$, $w_1 = 5$. The maximum processing time of a loop in the system $T_{max} = 1$ s and the maximum probability of system failure $P_{max} = 0{,}2$ are taken as constraints.

TABLE I. INITIAL DATA

| Device Type | $C_i$ | $N_i$ | $R_i$ | $P_i$ | $T_i$ | $M_i$ | $\tau_i$ |
|---|---|---|---|---|---|---|---|
| $u_1$ | 1000 | 0 | 512 | 0,01 | 0,01 | 4 | 0 |
| $u_2$ | 80 | 8 | 0 | 0,005 | 0 | 4 | 0,1 |

The $u_1$ processor can be, for example, a programmable logic controller, which has a fairly high cost, but does not have input-output channels. The $u_1$ repeater can be an input-output module, to which up to eight physical signals can be connected, or a network switch. Both types have network ports on board (4 pcs.) for connecting downstream devices.

The results of the program operation for different values of $A$ and $S$ are summarized in Table II.

TABLE II. RESULTS

| Number of loops | Number of levels $S$ | Cost | $T_{min}$ | $P_{sys}$ |
|---|---|---|---|---|
| 2 | 3 | 1160 | 0.12 | 0.0199 |
| 10 | 3 | 1240 | 0.2 | 0.0248 |
| 30 | 3 | 2400 | 0.25 | 0.0442 |
| 50 | 3 | 3640 | 0.28 | 0.0678 |
| 100 | 3 | 7360 | 0.31 | 0.1354 |
| 150 | 3 | - | - | - |
| 150 | 4 | 11000 | 0.38 | 0.1941 |

It is noteworthy that for $A = 150$ и $S = 3$ the algorithm did not find a feasible solution due to the existing limitations.

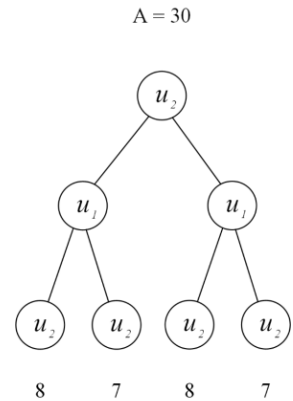

Fig. 2. Structure of the system for $A = 30$ and $S = 3$

A graphical representation of the optimal structures for some values of $A$ and $S$ is shown in Fig. 2, 3.

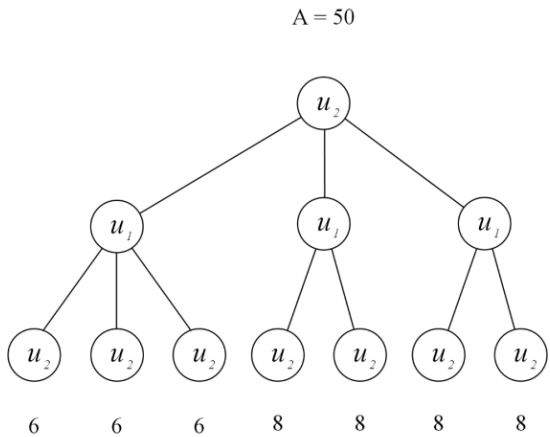

Fig. 3. Structure of the system for $A = 50$ and $S = 3$

The ant colony algorithm in its pure form is an approximate algorithm and does not provide an exact solution to the problem, however, based on this algorithm, it is possible to generate a reference solution, which can be improved by applying one of the known methods [6]. To improve the quality of the results, the ant colony algorithm can be modified by adding, for example, a local search. The applied algorithm has the property of getting stuck on local optima, which is especially pronounced with an increase in the dimensionality of the problem. To minimize the effect of getting stuck on local optima, it is necessary to select the parameters of the algorithm, or combine it with some additional method [15].

## VII. Further research

To obtain practically valuable results, the algorithm for solving the problem can be modified taking into account the recommendations given in the previous paragraph. It should be borne in mind that the solution to the problem is given under certain assumptions, such as the absence of horizontal links between hierarchy nodes, a predetermined partitioning of the array of input and output signals into circuits, the absence of delays in data transmission by processors, independence of failures of system nodes, etc. Taking into account these assumptions in the model can be the subject of additional studies that develop and generalize the results of this article.

The quality of the resulting control system can be verified by simulating the system with a specific control object, for example, a process plant of a chemical plant. It is noteworthy that the resulting system will be a control system with telecommunication channels [16], the analysis and synthesis of which are also the subjects of a separate study.

## VIII. Conclusion

The task of finding the optimal structure of a system built on the basis of components mass-produced by industry is extremely relevant. At present, the synthesis of the structure of such systems is performed mainly empirically based on the experience of the designer and recommendations of equipment manufacturers and is not always optimal. The proposed formalization of the problem and the solution method allow us to formulate an algorithm for synthesizing the optimal structure in terms of cost with restrictions on information capacity (number of channels), memory, speed and reliability.